\documentclass[12pt]{iopart}

\usepackage{iopams}
\usepackage{color}
\usepackage[english]{babel}
\usepackage{braket}

\usepackage[pdftex]{graphicx}

\definecolor{myblue}{rgb}{0.2,0.2,0.8}
\definecolor{myblack}{rgb}{0,0,0}

\usepackage[colorlinks=true,citecolor=myblue,linkcolor=myblack]{hyperref}

\def\ee{\mathord{\rm e}}
\def\ii{\mathord{\rm i}}

\newcommand{\rmP}{\mathrm{P}}
\newcommand{\dRenorm}{\tilde\Delta}
\newcommand{\tensor}[1]{\mathbf{#1}}

\newcommand{\eexp}[1]{\ee^{#1}}

\newcommand{\klamm}[1]{\mathrm{(#1)}}

\begin{document}

\title{ Two-Photon Scattering in USC regime}

\author{Vanessa Paulisch$^1$, Tao Shi$^2$, Juan Jos\'e Garc\'ia-Ripoll$^3$}

\address{$^1$ Max-Planck-Institute of Quantum Optics, Hans-Kopfermann-Strasse 1, 85748 Garching, Germany}
\address{$^2$ CAS Key Laboratory of Theoretical Physics, Institute of Theoretical Physics, Chinese Academy of Sciences, P.O. Box 2735, Beijing 100190, China}
\address{$^3$ Instituto de F{\'\i}sica Fundamental IFF-CSIC, Calle Serrano 113b, E-28006 Madrid, Spain}

\ead{vanessa.paulisch@mpq.mpg.de}

\begin{abstract}
  In this work we study the scattering of pairs of photons by a two-level system ultrastrongly coupled to a one-dimensional waveguide. We describe this problem using a spin-boson model with an Ohmic environment $J(\omega)=\pi\alpha\omega^1.$ We show that when coupling strength lays is about $\alpha\leq 1,$ the dynamics is well approximated by a polaron Hamiltonian, under the approximation of a conserved number of excitations. In this regime, we develop analytical predictions for the single- and two-photon scattering matrix computed with a Green's function method.
\end{abstract}


\vspace{2pc}

 
\maketitle

\section{Introduction}

A quantum emitter interacting with a photonic waveguide is classified as \emph{weak, strong or ultrastrongly} coupled, depending on the emitter resonance $\Delta,$ the rate of spontaneous emission $\Gamma,$ and the strength of non-radiative losses $\gamma.$ State-of-the-art waveguide-QED experiments work in the \emph{strong coupling regime} $\gamma \ll \Gamma \ll \Delta.$ These experiments are described with rotating wave approximation (RWA) methods: wavefunctions\ \cite{shen05,shen07a}, input-output-theory\ \cite{fan10,caneva15}, path integral formalism\ \cite{shi09,shi15} or diagrammatic approaches\ \cite{pletyukhov12,hurst17,kocabas16}. This description breaks down for \emph{ultrastrong coupling} (USC) experiments, a regime of broadband interactions that approach the speed of emitter oscillations $\Delta \sim \Gamma \gg \gamma.$ The USC regime in cavities\ \cite{niemczyk10,forn-diaz10} is fully understood, because it can be easily simulated and admits a full analytical description\ \cite{braak11}. However, a complete study of waveguide-QED USC experiments\ \cite{forn-diaz16} is still an open problem: since the RWA does not apply, a rigorous description demands costly simulations\ \cite{peropadre13, bera16, gheeraert17} based on numerical renormalization group or matrix product state (MPS) techniques. Therefore, it is desirable to replace those simulations with an analytical theory that delivers accurate predictions and a good intuition of the USC regime.

In this work we show that the approximation techniques from Shi\ \emph{et al.} \cite{shi17} provide a full, analytically tractable description of USC experiments with one and two photons. In section\ \ref{sec:Polaron} we extend the method by Shi\ \emph{et al.} to create an excitation conserving Hamiltonian for USC waveguide-QED experiments. In section\ \ref{sec:Analytics} we show how to use this Hamiltonian to compute the scattering matrix of one and two photons using the Green's function formalism, as further spelled out in \ref{app:Analytics}. In section\ \ref{sec:Numerics} we verify the approximations for one and two photons numerically. Using MPS simulations of the full spin-boson, we show that the conservation of excitations holds for a broad range of interactions up to $\Gamma/\Delta \simeq 40\%.$ In this regime, the approximate two-photon Hamiltonian describes the dynamics and can be simulated using a simple wavefunction method. We close this work with a summary of results and outlook for future work in section\ \ref{sec:summary}.

\section{Light-matter interaction with the polaron Hamiltonian}\label{sec:Polaron}

The spin-boson model is a good description for a two-level quantum emitter interacting with a bath of propagating photons as in waveguide QED experiments. The model usually takes the form
\begin{equation}
H
	= \frac{\Delta}{2} \sigma^z 
		+ \sum_k \omega_k a_k^\dagger a_k
		+ \sum_k g_k \sigma^x \left( a_k^\dagger + a_k \right).
\end{equation}
The constant $\Delta$ is the qubit resonance or gap. There is a bath of photons labeled by quasimomentum $k$, with frequencies $\omega_k$, that are anihilated and created by the Fock operators $a_k$ and $a_k^\dagger.$ The emitter interacts with the photons through the coupling constants $g_k.$ These are usually combined to form the spectral function, which is approximately Ohmic for many waveguides of interest
\begin{equation}
  J(\omega) 
  = 2 \pi \sum_k |g_k|^2 \delta(\omega - \omega_k)
  = \pi \alpha \omega^1.
\end{equation}
When $\alpha$ is very small, the spontaneous emission rate of the quantum emitter is given by $\Gamma\simeq J(\Delta)$. In other regimes, this is no longer true, and the two-level system experiences a large frequency renormalization and a large broadening of its resonance\ \cite{shi17}.

In most experiments, photons have a symmetric spectrum $\omega_{-k} = \omega_k$ and interaction with the quantum emitter $g_{-k} = g_k$. In this case, we can work with even and odd modes, $A_{k} = \frac{1}{\sqrt{2}} \left(a_k + a_{-k}\right)$ and $B_k = \frac{1}{\sqrt{2}} \left(a_k - a_{-k}\right)$. While the odd modes only evolve under the free Hamiltonian the even modes couple to the emitter with a coupling of $\tilde{g}_k=\sqrt{2} g_k$
\begin{equation}\label{eq:HamOriginal}
  H
  = \frac{\Delta}{2} \sigma^z 
  + \sum_{k\geq 0} \omega_k A_k^\dagger A_k
  + \sum_{k\geq 0} \omega_k B_k^\dagger B_k
  + \sum_{k\geq 0} \tilde{g}_k \sigma^x \left( A_k^\dagger + A_k \right).
\end{equation}
This means that the dynamics of the full waveguide is obtained from a free noninteracting problem with photons $B_k,$ and from a chiral scattering problem with one-directional modes $\{\sigma^z,A_k\}.$ In the rest of this work we will focus on this last problem, replacing $\tilde{g}_k$ with $g_k,$ and omitting the odd modes.

In Ref.\ \cite{shi17} we showed that it is convenient to transform the spin-boson Hamiltonian using a polaron transformation $H_\rmP = U_\rmP^\dagger H U_\rmP$ to make the ground state of the new operator close to a disentangled state $\ket{0,0}$. The transformation used is
\begin{equation}
U_\rmP = \exp \left[- \sigma^x \sum_k f_k (A_k^ \dagger - A_k )\right].
\end{equation}
The optimal displacements $f_k$ are obtained by minimizing the energy  $\braket{H_\rmP}$ over all product states.
This gives two self-consistent equations\ \cite{diaz-camacho16}
\begin{equation}
f_k 
	= \frac{g_k}{\omega_k + \dRenorm}, \qquad
\dRenorm
	= \Delta \exp \left[-2 \sum_k |f_k|^2 \right],
\end{equation}
which can be solved numerically. Up to constant energy shifts, the polaron Hamiltonian in the new basis reads
\begin{equation}\label{eq:HamPolaron}
H_P 
= \frac{\tilde{\Delta}}{2} \sigma^z O_{-\mathbf{f}}^\dagger O_{\mathbf{f}}
+ \sum_k \omega_k A_k^\dagger A_k
+ \tilde{\Delta} \sum_k f_k \sigma^x (A_k^\dagger + A_k).
\end{equation}
The parameter $\dRenorm$ is identified as the renormalized two-level gap; the operator $O_{\mathbf{f}} = \exp \left[ 2 \sigma^x \sum_k f_k A_k \right]$ (see \ref{app:Polaron}) is a normal ordered source of quantum fluctuations and the constants $f_k$ play the role of the renormalized coupling strengths.

At first glance, this might not seem like an improvement, as the new expression is still analytically untractable. However, from numerical simulations we know that the number of excitations $N=\frac{1}{2}(\sigma^z+1)+\sum_k A_k^\dagger A_k$ is approximately conserved. We therefore feel empowered to project the polaron model onto a sector with a fixed value of $N.$ To do this, we begin by introducing $F = \sum f_k A_k$ and expanding in powers of the interaction strength $\alpha$
\begin{equation}
  O_{\mathbf{f}} = 1 + 2 \sigma^x F + 2 F F + \mathcal{O}(\alpha^{3/2}).
\end{equation}
Up to second order, we recover a the Hamiltonian
\begin{eqnarray}\label{eq:HamPolaron2}
H_P^{(2)}  &=& H_0
              +\delta_0 \left( F^\dagger \sigma^- + \sigma^+ F \right)
              - \delta_0 \sigma^z F^\dagger F
              + \delta_0 \sigma^z (F F + \mathrm{h.c.})  \\
            && - \delta_0
              (\sigma^z \sigma^x F^\dagger F F + \mathrm{h.c.})
              + \delta_0 \sigma^z F^\dagger F^\dagger F F \nonumber
\end{eqnarray}
with the free part
\begin{eqnarray}
H_0 = \sum_k \omega_k A_k^\dagger A_k + \frac{\tilde{\Delta}}{2}\sigma^z,
\end{eqnarray}
and the constant $\delta_0 = 2 \tilde{\Delta}$. From this expansion, it is clear that the Hamiltonian in the single excitation subspace conserves the number of excitations
\begin{equation}
H_P^{(1)} 
	= H_0 
		+\delta_0\left( F^\dagger \sigma^- + \sigma^+ F \right)
		+\delta_0 F^\dagger F.
\end{equation}
For higher order terms we apply the RWA, neglecting the terms $\sigma^z F F$ and $\sigma^- F^\dagger F F.$
\begin{eqnarray}
\label{eq:HamPolaron2RWA}
H_\mathrm{RWA}^{(2)}  &=& H_0
              +\delta_0 \left( F^\dagger \sigma^- + \sigma^+ F \right)
              - \delta_0 \sigma^z F^\dagger F
              + \delta_0 \sigma^z (F F + \mathrm{h.c.})  \\
            && - \delta_0
              (\sigma^z \sigma^x F^\dagger F F + \mathrm{h.c.})
              + \delta_0 \sigma^z F^\dagger F^\dagger F F. \nonumber
\end{eqnarray}
Note, that the RWA can be applied here for larger coupling strengths $\alpha$ because the couplings to states with different excitation numbers scales with $\alpha$ instead of $\sqrt{\alpha}$ in the original frame.

\section{Two-photon scattering matrix}\label{sec:Analytics}

The photon-qubit scattering has been thourougly studied in recent years for waveguides with linear dispersion\ \cite{shen07,shi09,fan10,pletyukhov12,xu15,caneva15,hurst17,shi15} and also for dispersive waveguides\ \cite{shi09,zhou08,roy11,kocabas16}. Unfortunately, these results can only be partly generalized to the Hamiltonian\ (\ref{eq:HamPolaron2RWA}), due to the coupling $\delta_0 f_k$ and the USC corrections $F^\dagger F,$ or $\sigma^+F^\dagger F^2.$ In order to go beyond the predictions from Ref.\ \cite{shi17}, we need to develop a more sophisticated scattering method. We begin in section\ \ref{sec:1PhScattering} by rederiving the single-photon scattering matrix using the Green's function $G^\klamm{1}$ in the one excitation subspace. The results agree with earlier predictions, and can be used as a foundation for the two-photon scattering matrix, derived in section\ \ref{sec:2PhScattering}. Note that in order to simplify the presentation, all theoretical predictions are particularized to a linear dispersion $\omega_k = k$ with an exponential cutoff $g_k = \sqrt{\pi \alpha \omega} \eexp{-\omega / 2 \omega_c}$.

\subsection{Single Photon Scattering}\label{sec:1PhScattering}

As in Ref.\ \cite{shi17}, we study the scattering between assymptotically free states, from an initial state $\ket{\psi_i} $ to a final state $\ket{\psi_f}$. The scattering amplitude between these two assymptotically free states is defined as
\begin{equation} \label{eq:Smat}
S_{f;i} 
	= \lim\limits_{t_{f,i} \rightarrow \pm \infty}
		\eexp{\ii E_f t_f} \bra{\psi_f} \eexp{-\ii H (t_f - t_i)} \ket{\psi_i} \eexp{-\ii E_i t_i},
\end{equation}
where the Hamiltonian is $H = H_\rmP^\klamm{1}$ and the energies $E_{i/f}$ are the respective eigenenergies far away from the emitter, i.e., $E_{f/i} \ket{\psi_{f/i}} = H_0 \ket{\psi_{f/i}}$. For finding an analytical expression for the scattering amplitude, it turns out to be useful transform into an interaction picture with respect to $H_0$. In this rotated frame, the scattering amplitude,
\begin{equation}
S_{f;i} 
= \bra{\psi_f} \exp \left[-\ii \int_{-\infty}^{\infty} \mathrm{d}t\ V^{(1)}(t) \right] \ket{\psi_i},
\end{equation}
only depends on the interaction term in the interaction picture. This interaction term can be written as
\begin{equation}
V^{(1)} 
	= H_P^\klamm{1} - H_0
	= \vec{O}_1^\dagger \mathbf{u}_1 \vec{O}_1,
\end{equation}
with a vector of operators $\vec{O}_1 = (b, F)$ and an interaction matrix $\mathbf{u}_1 = (0, \delta_0 ; \delta_0 , -\delta_1)$.

This expression can be further evaluated by expanding the exponential in a Dyson series and using the fact that there is at most one excitation in the system, so that one can insert projectors onto the ground state in between the creation and annihilation operators in the interaction term, i.e., $V^\klamm{1} = \vec{O}_1^\dagger \ket{0} \mathbf{u}_1 \bra{0} \vec{O}_1 $. When summing all orders of the Dyson expansion (see \ref{app:Dyson}), one finds that the scattering amplitude
\begin{equation}
S_{f;i} 
	= \langle \psi_f | \psi_i \rangle 
		- 2 \pi \ii \delta(E_f - E_i) 
		\bra{\psi_f} \vec{O}_1^\dagger \ket{0} \mathbf{T}^\klamm{1} (E_i)
		\bra{0} \vec{O}_1 \ket{\psi_i},
\end{equation}
separates into a non-interacting part and a scattered part defined by a $\mathbf{T}^\klamm{1}$-matrix. This matrix contains an infinite sum, that converges to
\begin{equation}
\mathbf{T}^\klamm{1} (z)
	= \mathbf{u}_1 
		\sum_{n=0}^\infty \left( \mathbf{\Pi}^\klamm{1} (z) \cdot \mathbf{u}_1 \right)^n 
	=\left[ \mathbf{u}_1^{-1}  - \mathbf{\Pi}^{(1)}(z) \right]^{-1},
\end{equation}
where we introduced the self energy bubble $\mathbf{\Pi}^\klamm{1}$. This matrix can be calculated from the Green's function $G^\klamm{0}$ as 
\begin{equation}
\mathbf{\Pi}^\klamm{1} (z)
	= - \ii \int_{0}^{\infty} \rmd t \bra{0} \vec{O}_1(t) \vec{O}_1^\dagger \ket{0} \eexp{\ii z t}
	= \mathrm{diag} \left(G^\klamm{0}_{bb}(z), G^\klamm{0}_{FF}(z) \right),
\end{equation}
where the elements of the diagonal self energy bubble are $G^\klamm{0}_{bb}(z) = (z-\tilde{\Delta})^{-1}$ and $G^\klamm{0}_{FF}(z) = \Sigma(z)/(4 \tilde{\Delta^2})$. The self energy
\begin{equation}
\Sigma(\omega) 
	= 4 \tilde{\Delta}^2 \sum_k \frac{|f_k|^2}{\omega - \omega_k + \ii \eta}
	\equiv \delta_L(\omega) - \ii \Gamma(\omega) / 2.
\end{equation}
contains a Lamb shift $\delta_L$ and a decay rate $\Gamma$, which determine the scattering characteristics.

By straightforwardedly inverting the two-by-two matrices, we can obtain the full $\mathbf{T}^\klamm{1}(z)$-matrix. However, for the scattering of a single photon, the only relevant contribution comes from 
\begin{equation}
\mathbf{T}^\klamm{1}_{22}(z)
	= \frac{4 \tilde{\Delta}^2 \chi(z)}{ (z -\tilde{\Delta} )- \chi(z) \Sigma(z)},
\end{equation}
where we have introduced the factor $\chi(z) = \frac{z + \tilde\Delta}{2\tilde{\Delta}}$. Due to the energy conservation term in the scattering amplitude, an initial photonic states $\ket{\psi_i} = A_k^\dagger \ket{0}$ can only scatter to a state $\ket{\psi_f} = s_k A_k^\dagger \ket{0}$ with $|s_k|=1$, where the chiral phase shift $s_k$ is given by
\begin{equation}
s_k 
	= 1 - \ii  \frac{f_k^2}{\omega'(k)} \mathbf{T}^\klamm{1}_{22}(\omega_k)
	= 1 - \ii \frac{ -\chi(\omega_k)  \Gamma(\omega_k) }{(\omega_k- \tilde{\Delta})- \chi(\omega_k) \Sigma(\omega_k)}
	= \frac{h(\omega_k)^*}{h(\omega_k)}
\end{equation}
with $h(\omega) = (\omega- \tilde{\Delta})- \chi(\omega) \Sigma(\omega)$. To obtain this expression, we used the relation $4 \tilde{\Delta}^2 f_k^2 / \omega'(\omega_k) = \Gamma(\omega_k) $. From the chiral phase shift one can calculate the transmission and reflection coefficents as $t_k = \frac{1}{2} (s_k + 1)$ and $r_k = \frac{1}{2} (s_k - 1)$.

When the effective photon-photon interaction term $F^\dagger F$ can be neglected, we find $\chi(z) = 1$. In the standard case, where the Lamb shift vanishes $\delta_L \sim 0$, the decacy rate is uniform $\Gamma(\omega) \sim \Gamma$, we then recover the standard result with 
\begin{equation}
r_k = \frac{-\rmi \Gamma/2}{(\omega_k - \tilde{\Delta}) + \rmi \Gamma / 2}.
\end{equation}

\subsection{Two Photon Scattering}\label{sec:2PhScattering}

We now focus on finding an analytical expression for the scattering matrix of two photons scattering on a single emitter in the polaron frame. We solve the two photon scattering by using the hardcore boson representation\ \cite{batyev84} with annihilation (creation) operators $b$ ($b^\dagger$). In this representation, the spin operators are replaced by $\sigma^z \rightarrow 2 b^\dagger b -1$ and $\sigma^- \rightarrow b$. To recover the results for a two level system, one has to introduce an energy penalty for double excitations, $u_0 b^\dagger b^\dagger b b$, which has to be taken to $u_0 \rightarrow \infty$ at the end of the calculations. Under these considerations, the Hamiltonian we aim to approximate is
\begin{equation}
H_\rmP^\klamm{2}
	=  H_\rmP^{(1)} 
	- 2 \delta_0 b^\dagger F^\dagger b F
	- \delta_0
	(b^\dagger F^\dagger F F + \mathrm{h.c.})
	- \delta_0 F^\dagger F^\dagger F F
	+ u_0 b^\dagger b^\dagger b b.
\end{equation}

The scattering matrix between two assymptotically free states, from $\ket{\psi_{i}}$ to $\ket{\psi_{f}}$, with eigenenergies $H_0 \ket{\psi_{f/i}} = E_{f/i} \ket{\psi_{f/i}}$ is given by Equation \ref{eq:Smat} where the Hamiltonian this time is $H = H_\rmP^\klamm{2}$. Instead of transforming to the interaction picture with respect to the free Hamiltonian $H_0$, we use a frame rotating with the single excitation Hamiltonian $H_\rmP^\klamm{1}$. This transformation allows us to relate the two photon scattering amplitude to expressions obtained in the single excitation subspace, which we solved analytically in section \ref{sec:1PhScattering}. The scattering amplitude can then be calculated using a Dyson expansion of
\begin{equation}
S_{f;i} 
	= \lim\limits_{t_{f/i} \rightarrow \pm \infty}
		\eexp{\ii E_f t_f} \bra{\psi_f (t_f)}
		\exp \left[-\ii \int_{-\infty}^{\infty} \mathrm{d}t\ V^\klamm{2}(t) \right] 
		\ket{\psi_i (t_i)} \eexp{-\ii E_i t_i},
\end{equation}
where $\ket{\psi_{f/i}(t_{f/i})} = \exp\left[\ii H_\rmP^\klamm{1} t_{f/i}\right] \ket{\psi_{f/i}} $. Similarly to the expression in the single excitation regime, the interaction term can be written as a product of vectors and matrices, 
\begin{equation}
V^\klamm{2}
	= H_P^\klamm{2} - H_P^\klamm{1}
	= \vec{O}_2^\dagger \mathbf{u}_2 \vec{O}_2,
\end{equation}
where the interaction matrix $\mathbf{u}_2$ is now a three-by-three matrix and the vector of operators $\vec{O}_2 = ( bb, bF, FF)$   now has to be taken in the interaction picture rotating with $H_\rmP^\klamm{1}$.
As there are two excitations in the system at all times, we can project onto the vacuum in between the vectors of the two creation and anihilation operatorse, that is, we can write $V^\klamm{2} = \vec{O}_2^\dagger \ket{0}\mathbf{u}_2 
\bra{0} \vec{O}_2$.

As for the single photon case, the scattering matrix splits up into two parts when performing the Dyson expansion, i.e.,
\begin{equation}
S_{f;i} 
	= S_{f;i}^\mathrm{unco} 
		- 2 \pi i \delta(E_f - E_i) \vec{v}_f^\dagger \mathbf{T}^\klamm{2}(E_i) \vec{v}_i,
\end{equation}
where the uncorrelated part for an initial state $\ket{\psi_i} = A_{k_1}^\dagger A_{k_1}^\dagger \ket{0}$ to a final state $\ket{\psi_f} = A_{p_1}^\dagger A_{p_1}^\dagger \ket{0}$ is given by
\begin{equation}
S_{f;i}^\mathrm{unco}
	= s_{k_1} s_{k_2} 
		\left(\delta_{p_1 k_1} \delta_{p_2 k_2} + \delta_{p_2 k_1} \delta_{p_1 k_2} \right),
\end{equation}
and
\begin{equation}
  \label{eq:1}
  \vec{v}_{f/i}=\bra{0} \vec{O}_2 \ket{\psi_{f/i}}.
\end{equation}
If the two photons don't overlap in space or have a too narrow bandwidth, then this uncorrelated term is the only relevant one.

When the photonic modes overlap, the correlated part of the scattering is encoded in the $\mathbf{T}^\klamm{2}$-matrix. Just as for the single photon case, it can be obtained from the infinite sum
\begin{equation}
\mathbf{T}^\klamm{2} (z)
	= \mathbf{u}_2
		\sum_{n=0}^\infty \left( \mathbf{\Pi}^\klamm{2} (z) \cdot \mathbf{u}_2 \right)^n 
	=\left[ \mathbf{u}_2^{-1}  - \mathbf{\Pi}^{(2)}(z) \right]^{-1},
\end{equation}
where the self energy bubble $\mathbf{\Pi}^\klamm{2} (z)$ contains correlators of the four operators appearing in every combination of $(bb, bF, FF)$ as
\begin{equation}
\mathbf{\Pi}^\klamm{2} (z)
	= - \ii \int_{0}^{\infty} \rmd t \bra{0} \vec{O}_2(t) \vec{O}_2^\dagger \ket{0} \eexp{\ii z t}.
\end{equation}
By applying Wick's theorem\ \cite{wick50} one can reduce the four-point correlators to two two-point correlators, which can be expressed in terms of the Green's function $G^\klamm{1}$ in the single excitation subspace. In fact, one needs to calculate convolutions over these Green's Functions, for example for the element 
\begin{equation}
\mathbf{\Pi}^\klamm{2}_{11} (z) 
	= 2 \ii \int_{-\infty}^{\infty} \frac{\rmd \omega}{2 \pi} 
			G^\klamm{1}_{bb} (\omega) G^\klamm{1}_{bb} (z-\omega).
\end{equation}
For the ohmic model, the convolution can be calculated easily at least numerically, because $G^\klamm{1}_{bb}(z) = \bra{0} b \frac{1}{z - H_\rmP^\klamm{1}} b^\dagger \ket{0}$ can be obtained analytically (see \ref{app:Subspace1}) and doesn't have any poles along the real axis for $\alpha <0.5$.

For the final step, we note that the energy conservation for the scattering of a photonic initial state $\ket{\psi_i} = A_{k_1}^\dagger A_{k_2}^\dagger \ket{0}$ (and similarly for the final state) originates from the time dependence of  
\begin{equation}
\bra{0} \vec{O}_2(t) \ket{\psi_{i}} 
	= \vec{v}_{i} \frac{\delta_0 f_{k_1}}{h(\omega_{k_1})} \frac{\delta_0 f_{k_2}}{h(\omega_{k_2})} \eexp{-\ii (\omega_{k_1} + \omega_{k_2}) t},
\end{equation}
which at the same time defines the prefactor $\vec{v}_{f/i}$. This relation can shown by applying Wick's theorem and expresssing the two two-point correlators in terms of the Green's Function $G^\klamm{1}$.

By combining the above results, we obtain the correlated part of the scattering amplitude
\begin{equation}
S^\mathrm{corr}_{p_1 p_2; k_1 k_2} 
	= - 2 \pi \rmi \delta(E_f - E_i)
	\frac{\delta_0 f_{p_1}}{h(\omega_{p_1})} \frac{\delta_0 f_{p_2}}{h(\omega_{p_2})} 
	\vec{v_f}^\dagger \mathbf{T}^\klamm{2}(E_i) \vec{v_i}
	\frac{\delta_0 f_{k_1}}{h(\omega_{k_1})} \frac{\delta_0 f_{k_2}}{h(\omega_{k_2})},
\end{equation}
where $E_i = \omega_{k_1} + \omega_{k_2}$ and $E_f = \omega_{p_1} + \omega_{p_2}$. We have verified these results using the more formal and general method of path integrals.

For the standard result, where interaction terms containing more than one photon operator can be neglected, the only relevant term in the $\mathbf{\Pi}^\klamm{2}$-matrix is the element $\mathbf{\Pi}^\klamm{2}_{11}(z)$. For $\delta_L \sim 0$ and $\delta_0 f_k \sim \sqrt{\Gamma} $ the middle part is then given by $\vec{v_f}^\dagger \mathbf{T}^\klamm{2}(E_i) \vec{v_i} = 4 \left( z - 2 \tilde{\Delta} + \rmi \Gamma \right)$ so that we obtain the standard result\ \cite{fan10}
\begin{equation}
S^\mathrm{corr}_{p_1 p_2; k_1 k_2} 
	= \delta_{p_1 + p_2, k_1 + k_2}
	r_{p_1} r_{p_2}
	\left( r_{k_1} + r_{k_2} \right).
\end{equation}

\section{Numerical simulations}\label{sec:Numerics}

In this section we discuss the number conserving approximation\  (\ref{eq:HamPolaron2RWA}), comparing numerical simulations of this model with the exact polaron Hamiltonian\ (\ref{eq:HamPolaron}). We focus on scattering experiments with two photons, created on top of the ground state $\ket{gs}$
\begin{eqnarray}
  \label{eq:one-photon-state}
  \ket{\psi_1(0)} &=& \sum_{k} \phi(\omega_{k}, \mu_1, s_1, x_1)A_{k}^\dagger \ket{gs},\\
  \label{eq:two-photon-state}
  \ket{\psi_2(0)} &=& \sum_{k,p} \phi(\omega_{k}, \mu_1, s_1, x_1)\phi(\omega_{p}, \mu_2, s_2, x_2)A_{k}^\dagger A_{p}\ket{gs}
\end{eqnarray}
as Gaussian wavepackets centered on frequency $\mu$, position $x$ and width $s$
\begin{equation}
  \phi(\omega,\mu,x) = \frac{1}{\mathcal{N}^{1/2}} \exp\left( -\frac{(\omega - \mu)^2}{2 s} - \ii \omega x \right).
\end{equation}
For the numerical simulations, we use a discrete chiral model in which a waveguide of length $L$ is divided into $2 N + 1$ segments, with $N$ distinct modes $a_k$, a sine dispersion relation $\omega_k = \omega_c \sin\left( c k / \omega_c \right)$ and a hard frequency cutoff $\omega_c$. The Ohmic environment at low frequencies is simulated by the light-matter interaction $g_k = \sqrt{\pi \alpha \omega_k/L}.$

We study the evolution of this state using two different methods. We begin working with the number conserving Hamiltonian\ (\ref{eq:HamPolaron2RWA}). This operator is expanded on a basis of states with up to two excitations. The wavefunction spreads over a total of $1 + 2N + N(N-1)/2$ different states. We solve the Schr\"odinger equation using Chebyshev polynomials with up to $N=256$ modes. The outcome of these simulations is identical to the predictions from section\ \ref{sec:Analytics}, although we cannot use exactly the same formulas because of the slightly different spectrum and coupling constants.

The wavefunction simulations are compared with a matrix product state simulation of the polaron Hamiltonian\ (\ref{eq:HamPolaron}) without approximations. In order to avoid the long-range interaction term, which slows down simulations, we rewrite the model as a tight-binding Hamiltonian
\begin{eqnarray}
  H_P &=& \frac{\tilde\Delta}{2} \sigma^z e^{-2\theta\sigma^x c_0^\dagger} e^{+2\theta\sigma^x c_0/\beta_0}
  + \tilde{\Delta}\beta_0\sigma^x(c_0 + c_0^\dagger)\\
  &+& \sum_{r=0}^{N-1} \alpha_r c_r^\dagger c_r
  + \sum_{r=0}^{N-1} \beta_r (b_{r+1}^\dagger b_r + \mathrm{H.c.}) .\nonumber
\end{eqnarray}
This new form is exactly constructed using a Lanczos recursion\ \cite{chin10,zueco18}, starting from the Fock operator
\begin{equation}
  c_0 = \frac{1}{\theta} \sum_k f_k A_k,\;\theta^2 = \sum_k |f_k|^2,\;\beta_0=\tilde{\Delta}\theta
\end{equation}
and building the other operators so that $[c_r,H_P]=\alpha_r c_r + \beta_r c_{r+1} + \beta_{r-1} c_{r-1}.$ This new formulation of $H_P$ is a nearest-neighbor interaction Hamiltonian which allows a fast numerical simulation using MPS and Trotter expansions\ \cite{ripoll06}.

We have simulated the evolution of the one-photon\ (\ref{eq:one-photon-state}) and two-photon states\ (\ref{eq:two-photon-state}) using both methods. In order to compare simulations we have computed the excited state probability and number of excitations
\begin{eqnarray}
  P_e &=&\frac{1}{2} (\braket{\sigma^z}+ 1),\mbox{ and } \label{eq:Pe}\\
  N_{excit} &=&\frac{1}{2} (\braket{\sigma^z}+ 1) + \sum_k \braket{A_k^\dagger A_k}. \label{eq:Nexcit}
\end{eqnarray}
We also reconstructed the one- and two-photon components as
\begin{eqnarray}
  \Psi_1(\omega_k, t) &=& \braket{gs| A_k| \psi(t)},\\
  \Psi_2(\omega_k,\omega_p, t) &=& \braket{gs| A_kA_p|\psi(t)}.
\end{eqnarray}

\begin{figure}[t]
  \centering
  \includegraphics[width=\linewidth]{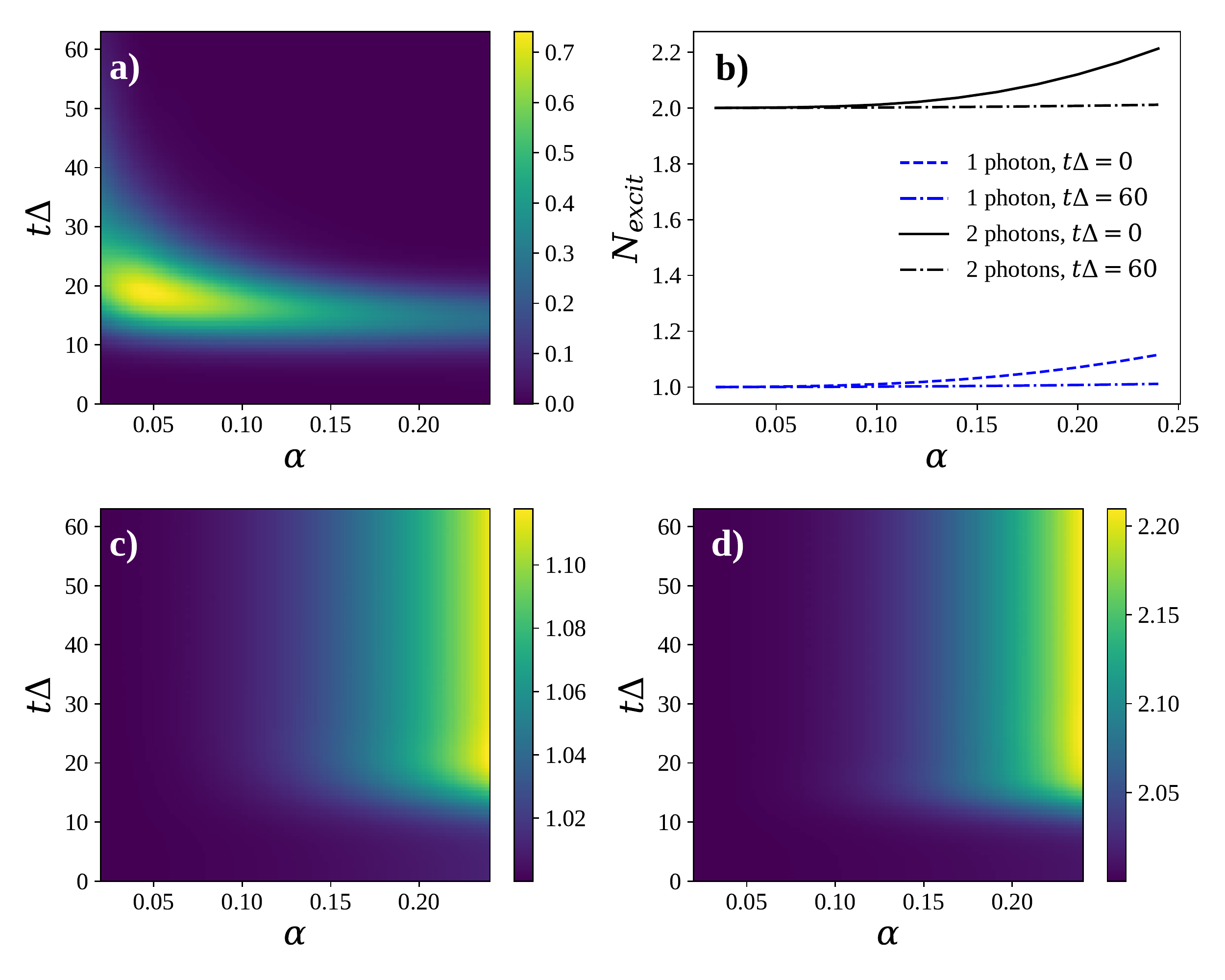}
  \caption{Number of excitations in one- and two-photon scattering. Figure (a) shows the excitation probability of the qubit $P_e$\ (\ref{eq:Pe}) as it interacts with the incoming photons. Figure (b) shows the asymptotic number of excitations $N_{excit}$\ (\ref{eq:Nexcit}), at a time $t\Delta\simeq 60$ where the emitter has relaxed, comparing it with the initial number of excitations at $t=0.$ We also plot density plots of the (c) one-photon and (d) two-photon scattering. The simulation was performed using MPS with $N=256$ modes, a bond dimension $\chi=80,$ and a cut-off $\omega_c=4\Delta.$}
  \label{fig:nphotons}
\end{figure}

We begin our study by analyzing the degree of our number conserving approximation, studying the number of excitations $N_{excit}$ as a function of time for different coupling strengths. Figure\ \ref{fig:nphotons}a shows the dynamics of the quantum scatterer when interacting with two incoming photons. We use this plot to get an idea of the interaction time and when the photons can be considered ``free'' again. Figure\ \ref{fig:nphotons}b plots the number of excitations before and after the scattering for one- and two-photon scattering. The initial number of excitations is very close to 1 and to 2, because the polaron Hamiltonian does a good job at disentangling the qubit in the ground state. As time evolves, the number of excitations grows slightly and saturates ---cf. figures\ \ref{fig:nphotons}c-d---, converging to a value that deviates about 0.2\% in the interval $\alpha\in[0,0.1].$

\begin{figure}[t]
  \centering
  \includegraphics[width=\linewidth]{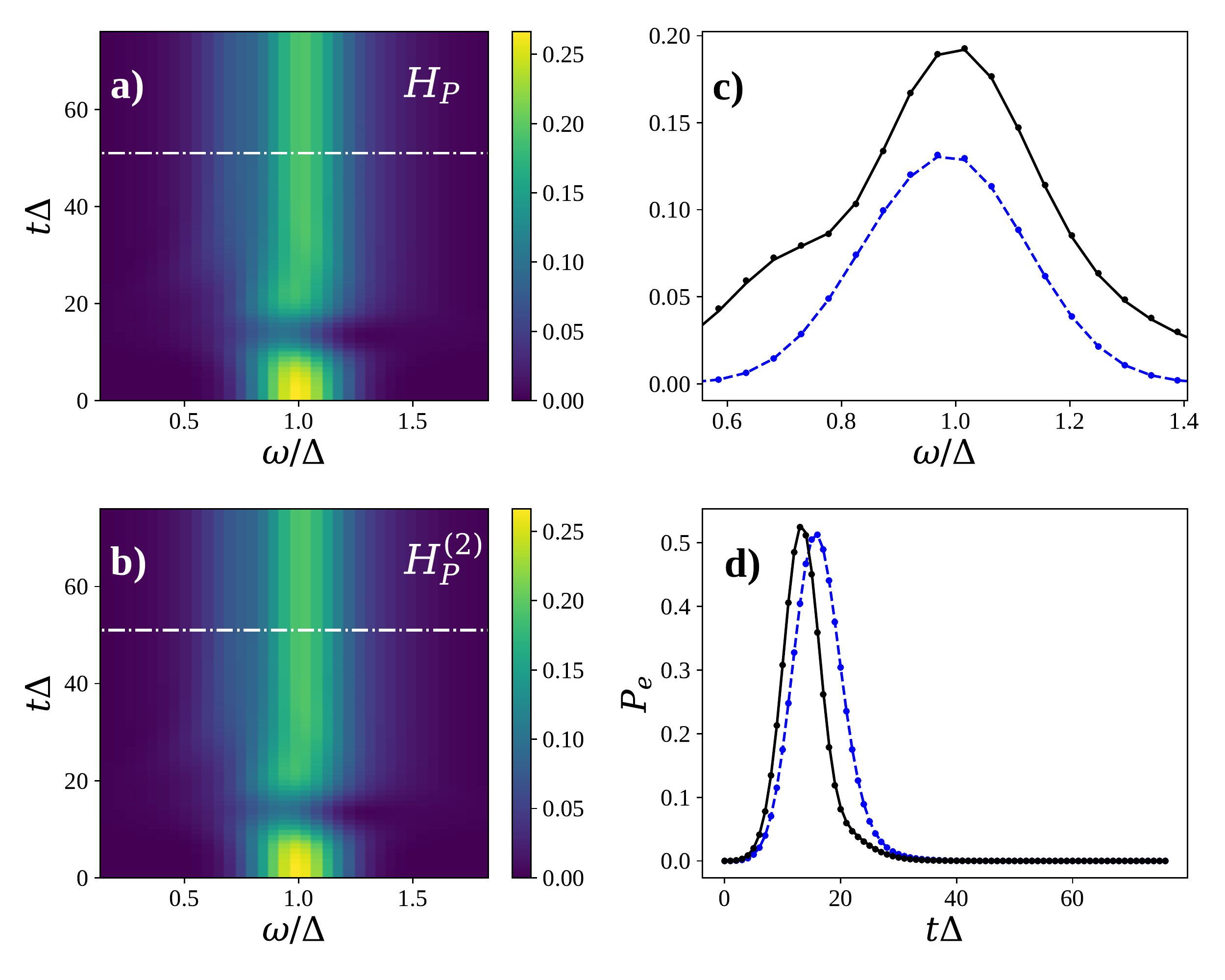}
  \caption{Scattering of one and two photons against a quantum emitter. Simulations with (a) the full Hamiltonian $H_P$ using MPS and (b) with the number conserving model $H_P^{(2)}$ using wavefunctions. In figures (a) and (b) we plot the integrated wavefunction $F(\omega,t) = \sum_{\omega_2} |\Psi_2(\omega,\omega_2,t)|^2.$ Figure (c) shows a cut of the evolution at $t\Delta=50$, plotting $|\Psi_1(\omega,t)|^2$ (dashed) and $F(\omega,t)$ (solid) for the full model, together with the approximations from $H_P^{(2)}$ (dots). Figure (d) shows the excited state probability $P_e$ for a single-photon (dashed) and two-photon (solid) scattering experiment, together with the approximate solution based on $H_P^{(2)}.$ Simulation parameters similar to figure\ \ref{fig:nphotons}.}
  \label{fig:scattering}
\end{figure}

Since the number of excitations is approximately conserved, we expect that the MPS simulations be well approximated by the simpler model~(\ref{eq:HamPolaron2RWA}). Figures\ \ref{fig:scattering}a-b compare both methods in a two-photon scattering experiment with $N=256$ modes, a cut-off $\omega_c=4\Delta$ and a coupling strength $\alpha=0.12.$ The numerical simulations with MPS were performed using a maximum bond dimension $\chi=80,$ integrating with third order Trotter method and a time-step of $0.05/\Delta.$ The left panels show a density plot of the integrated probability distribution over the second mode
\begin{equation}
  F(\omega,t) = \sum_{\omega_2}|\Psi_2(\omega,\omega_2)|^2,
\end{equation}
as a function of time. The simulation with MPS and the full model (figure\ \ref{fig:scattering}a) is indistinguishable from the approximate number conserving Hamiltonian (figure\ \ref{fig:scattering}b). Figure\ \ref{fig:scattering}c shows a transverse cut of the density plots. The solid line represents $F(\omega,t)$, computed with MPS at a time $t\Delta=50,$ well after the light-matter interaction finished. The line is extremely close to its approximation using $H_P^{(2)},$ shown as dots. For completeness, we also plot $|\Psi_1(\omega,t)|^2$ for a single-photon scattering experiment with both models. Finally, in figure\ \ref{fig:scattering}d we plot the reaction of the emitter to the incoming light $P_e(t),$ both for one (dashed) and two photons (solid), together with the number conserving approximation.
 
\begin{figure}[t]
  \centering
  \includegraphics[width=\linewidth]{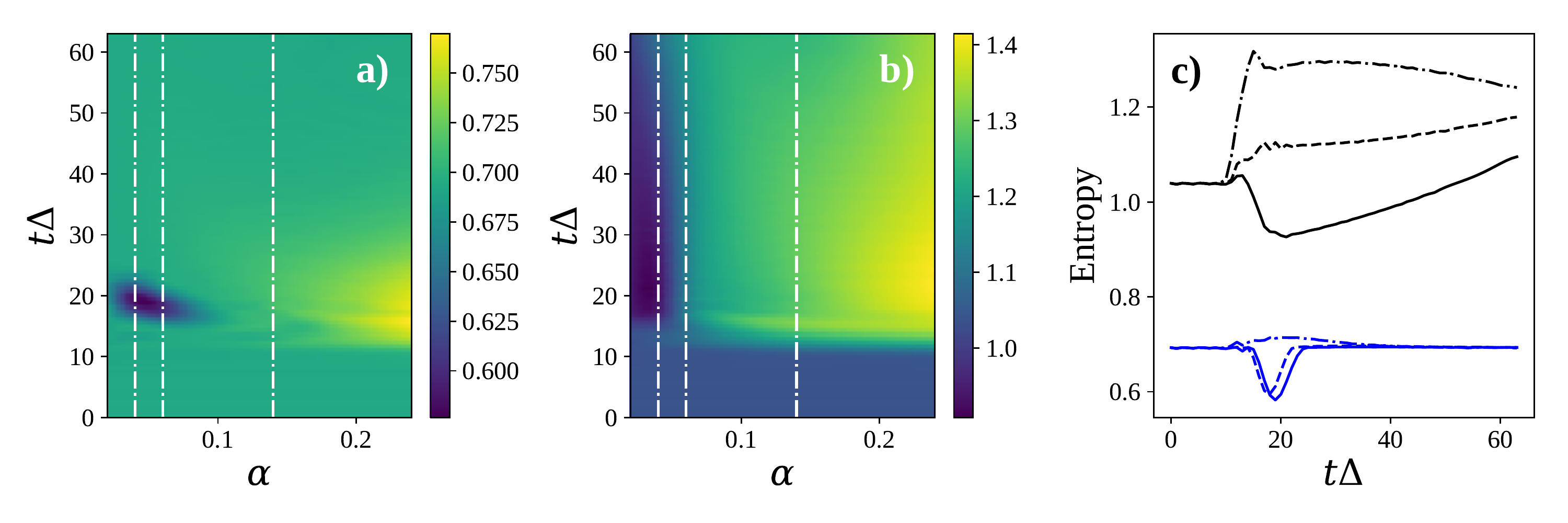}
  \caption{Maximum entanglement entropy in the MPS simulation of (a) one-photon and (b) two-photon scattering. Figure (c) shows vertical cuts of the previous plots for one- (bottom lnes) and two-photon scattering (top lines), at $\alpha=0.04, 0.06$ and $0.14$. Simulation parameters similar to figure\ \ref{fig:nphotons}.}
  \label{fig:entropy}
\end{figure}

We have also verified the numerical complexity and convergence of the MPS simulations using different bond dimensions and problem sizes. As part of this process we have computed the von Neumann entropy of the MPS wavefunction, with respect to bipartitions in $n+m=M$ modes. The entropy is a measure of how complex the state is and how bad the MPS approximation becomes. In our simulations, with up to 256 modes, this entropy converges already for small bond dimensions $\xi \sim 60.$

The study of the entropy also reveals differences between the single-photon and two-photon scattering. As we see in figure\ \ref{fig:entropy}a, entropy decreases when one photon is absorbed and recovers to the original value once it is reemitted. In the two-photon case, shown in figure\ \ref{fig:entropy}b, the dip can be almost imperceptible, because it is overwhelmed by the entanglement between the absorbed photon and the photon that continues travelling. This said, figure\ \ref{fig:entropy}c shows that even in this case the total entanglement remains within acceptable bounds in our simulations.

\section{Summary and discussion}
\label{sec:summary}

Summing up, in this work we have developed a scattering theory for one and two photons interacting with a two-level system in a one-dimensional waveguide. The theory extends a previous method to develop number-conserving RWA Hamiltonians that describe the light-matter interaction in the strong- and ultrastrong coupling regimes\ \cite{shi17}.

We have validated the predictions of this method, comparing with optimized MPS simulations using the exact polaron Hamiltonian in a tight-binding rewrite that minimizes the simulation complexity\ \cite{chin10,zueco18}. The simulations confirm the quantitative accuracy of our scattering theory for USC coupling strengths of up to $\alpha\simeq 0.1.$ Beyond this coupling we see a small fraction of excitations that are unaccounted for by our theory. These could be explained by recent works on entangled pairs production in the USC regime\ \cite{gheeraert18}, but this question requires a more complicated processing of the MPS state.

The predicitions in this manuscript can be verified in existing setups with superconducting qubits\ \cite{forn-diaz16}, with the help of tomographic methods that reconstruct the scattering matrix from homodyne measurements\ \cite{ramos17}.

J.J.G.R. acknowledges support from MINECO/FEDER Project FIS2015-70856-P, CAM PRICYT project QUITEMAD+CM S2013-ICE2801. T.S. acknowledges support by the ThousandYouth-Talent Program of China. VP acknowledges the Cluster of Excellence Nano Initiative Munich (NIM). J.J.G.R. thankfully acknowledges the computer resources provided by the Centro de Supercomputaci\'{o}n y Visualizaci{\'o}n de Madrid (CeSViMa) and the Spanish Supercomputing Network.

\appendix

\section{Polaron Transformation in Detail}\label{app:Polaron}

The polaron transformation $U_P = \exp \left[ - \sigma^x \sum f_k (A_k^\dagger - A_k) \right]$ acts on the operators as 
\begin{eqnarray}
U_P^\dagger A_k^\dagger U_P 
	=& A_k - \sigma^x f_k,\\
U_P^\dagger \sigma^x U_P 
	=& \sigma^x,\\
U_P^\dagger \sigma^z U_P 
	=& \sigma^z \eexp{-2 \sigma^x \sum_k f_k \left(A_k^\dagger - A_k\right)}
	=  \frac{\tilde{\Delta}}{\Delta} \sigma^z O_{-\mathbf{f}}^\dagger O_{\mathbf{f}},
\end{eqnarray}
The operators $O_\mathbf{f}$ are normal ordered exponentials
\begin{equation}
  O_{\mathbf{f}} = \exp \left[ 2 \sigma_x \sum_k f_k A_k \right],
\end{equation}
from which we have extracted a renormalization factor that gets inserted into the emitter's frequency $\tilde{\Delta} = \Delta \eexp{-2 \frac{1}{L} \sum_k |f_k|^2} $. The Hamiltonian in the polaron frame is then given by
\begin{equation}
H_P 
	= U_P^\dagger H U_P 
	= \frac{\tilde \Delta}{2} \sigma^z 
		+ \sum_k \omega_k A_k^\dagger A_k
			+\sum_k g_k \sigma^x \left( A_k^\dagger + A_k \right)
			+ E_0
\end{equation}
with the energy shift $E_0 = \sum_k \omega_k |f_k|^2 - 2 g_k f_k $.

Minimizing the energy of the vacuum state $E_0[f_k]=\braket{0|H_P|0} = - \tilde{\Delta}/2 + E_0$ over all variational parameters $\{f_k\}$ we obtain a self-consistent equation
\begin{equation}
  f_k 
  = \frac{g_k}{\omega_k + \dRenorm}, \qquad
  \dRenorm
  = \Delta \exp \left[-2 \sum_k |f_k|^2 \right],
\end{equation}
that relates $\tilde\Delta$ and $f_k.$ This equation can be solved iteratively, starting with an estimate $\tilde\Delta \leq \Delta,$ and repeatedly computing a new set of $\{f_k\}$ until convergence is achieved.

\section{Analytical Results}\label{app:Analytics}
\subsection{Dyson Expansion}\label{app:Dyson}

For both the single photon and two photon scattering amplitude, we encounter time evolutions, which we expand in a Dyson series. In particular, we need to calculate
\begin{equation}
S_{f;i} 
	= \bra{\psi_f} \exp \left[-\ii \int_{t_i}^{t_f} \mathrm{d} t\ V(t) \right] \ket{\psi_i}.
\end{equation}
When the interaction term can be written as a product of matrices, $V = \vec{O}^\dagger \tensor{u} \vec{O}$ and when the vector of operators $\vec{O}$ anihilates all excitations in $\ket{\psi_{i/f}}$, one can project onto the ground state in between. In particular, we can replace the interaction term by $V = \vec{O}^\dagger \ket{0} \tensor{u} \bra{0} \vec{O}$.

We then expand the scattering amplitude in a Dyson series as
\begin{eqnarray}
S_{f;i} 
	&= \bra{\psi_f} 1 - \rmi \int_{-\infty}^{\infty} \rmd t_1\ V(t)
			+ (-\rmi)^2 \int_{-\infty}^{\infty} \rmd t \int_{-\infty}^{t} \rmd t_2 \ V(t_1) V(t_2)
			+ \ldots \ket{\psi_i}
	 \nonumber \\
	&= \braket{\psi_f}{\psi_i}
		-\rmi \int_{-\infty}^{\infty} \rmd t\ \eexp{\ii (E_f-E_i) t} 
			\vec{v}_f^\dagger 
			\mathbf{u} 
			\vec{v}_i
	\nonumber	\\
		& \qquad + (-\rmi )^2 \int_{-\infty}^{\infty} \rmd t \int_{0}^{\infty} \rmd \tau \
		 \eexp{\ii (E_f-E_i) t} \vec{v}_f^\dagger 
		\mathbf{u} 
		\langle 0 | \vec{O}(\tau) \vec{0}^\dagger|0 \rangle \eexp{\rmi E_i \tau}
		\mathbf{u} 
		\vec{v}_i
		\nonumber \\
	&= \langle \psi_f | \psi_i \rangle 
			-2 \pi \ii \delta (E_f - E_i)
			\vec{v}_f^\dagger
			\left[ \mathbf{u}  
			+ \mathbf{u} \Pi^{(1)}(E_i) \mathbf{u} + \ldots \right] 
			\vec{v}_i 
	\nonumber \\		
	&= \langle \psi_f | \psi_i \rangle 
			-2 \pi \ii \delta (E_f - E_i)
			\vec{v}_f^\dagger
			\left[ \mathbf{u}^{-1} - \Pi(E_i) \right]^{-1}
			\vec{v}_i,
\end{eqnarray}
where we have used $\bra{0} \vec{O}(t) \ket{\psi_i} = \vec{v}_f \eexp{-\rmi E_i t}$ and have defined the self energy bubble $\tensor{\Pi}(z) \equiv  \int_{0}^{\infty} \rmd \tau \langle 0 | \vec{O}(\tau) \vec{0}^\dagger|0 \rangle \eexp{\rmi z \tau}$.

\subsection{Single Excitation Subspace}\label{app:Subspace1}
In the single excitation subspace we find for the matrix $\tensor{T}^\klamm{1} = \left[ \mathbf{u}^{-1} - \Pi(E_i) \right]^{-1}$ the expression
\begin{eqnarray}
\tensor{T}^\klamm{1}(z) 
	=\left[ \tensor{u}_1^{-1} - \tensor{\Pi}^\klamm{1}(z) \right]^{-1}
	=\frac{1}{h(z)}
		\left( \begin{array}{cc}\
			(z-\dRenorm) \Sigma(z) & \delta_0 (z-\tilde{\Delta})\\ 
			\delta_0 (z-\dRenorm) & \delta_0^2 \chi(z)
		\end{array} \right), 
\end{eqnarray}
where we have defined the denominator $h(z) \equiv z-\dRenorm - \chi(z) \Sigma(z) $ and $\chi(z) \equiv \frac{z + \dRenorm}{2 \dRenorm}$.

The self energy $\Sigma(\omega)$ can be calculated for different dispersion relations and cut-offs and in the limit of infinite cut-off one obtains for the linear and modulus sine dispersions the value
\begin{eqnarray}
\Sigma(\omega)
=& \frac{2 \alpha \tilde{\Delta}^2}{(\omega+\tilde{\Delta})^2}
\left( \omega \ln \frac{\omega}{\tilde{\Delta}} - \omega - \tilde{\Delta} 
- \ii \pi \omega \right), \mathrm{for}\ \omega>0, \\
\Sigma(\omega)
=& \frac{2 \alpha \tilde{\Delta}^2}{(\omega+\tilde{\Delta})^2}
\left( \omega \ln \frac{-\omega}{\tilde{\Delta}} - \omega - \tilde{\Delta} \right),
\mathrm{for}\ \omega<0.
\end{eqnarray}
For some values ($\omega = 0, - \tilde{\Delta}$), one has to consider the correct limit and obtains $\Sigma(0) = - 2 \alpha \tilde{\Delta}$, and $\Sigma(-\tilde{\Delta}) = - \alpha \tilde{\Delta}$.

We can calculated the Green's function for one excitation from the T-matrix by using the relation
\begin{equation}
G^\klamm{1} 
= G^\klamm{0} + G^\klamm{0} T^\klamm{1} G^\klamm{0},
\end{equation} 
which yields for the elements $G^\klamm{1}_{O_1 O_2} = \bra{0} O_1 G^\klamm{1} O_2^\dagger \ket{0}$ the 

\begin{eqnarray}
G^\klamm{1}_{bb} (\omega)
	&= \frac{1 + \Sigma(\omega) \delta_1 / \delta_0^2}{h(\omega)},\\
G^\klamm{1}_{b A_k} (\omega)
	&= \frac{1}{h(\omega)}	
	\frac{\delta_0 f_k}{\omega - \omega_k}
	= G^\klamm{1}_{A_k b} (\omega),\\
G^\klamm{1}_{A_p A_k} (\omega)
	&= \frac{\delta_{pk}}{\omega - \omega_k } +
	\frac{\delta_0 f_p}{\omega - \omega_p }
	\frac{\chi(\omega)}{h(\omega)} 
	\frac{\delta_0 f_k}{\omega - \omega_k } , \\
G^\klamm{1}_{b F}(\omega)
	&= \frac{\Sigma(\omega) / \delta_0 }{h(\omega)} 
	= G^\klamm{1}_{F b} (\omega),\\
G^\klamm{1}_{F F}(\omega)
	&= \frac{(\omega-\tilde{\Delta}) \Sigma(\omega) / \delta_0^2}{h(\omega)}.
\end{eqnarray}

\subsection{Two Excitation Subspace}\label{app:Subspace2}

For the \textbf{energy conservation}, we used the fact that $\bra{0} \vec{O}_2(t) \ket{\psi_i} = \vec{v}_i \eexp{-\ii E_i t} \ket{\psi_i}$. To prove this relation, we first use 
\begin{eqnarray}
\bra{0} b(t) A_k^\dagger \ket{0}
	=& \ii \int_{-\infty}^{\infty} \frac{\rmd \omega}{2 \pi} G^\klamm{1}_{b A_k} (\omega^+) 
		\eexp{-\ii \omega^+ t}
	= \frac{\delta_0 f_k}{h(\omega_k)} \eexp{\ii \omega_k t}
	\equiv \beta_k \eexp{\ii \omega_k t}, \\
\bra{0} F(t) A_k^\dagger \ket{0}
	=& \ii \int_{-\infty}^{\infty} \frac{\rmd \omega}{2 \pi}
		G^\klamm{1}_{F A_k} (\omega^+)\eexp{-\ii \omega^+ t}
	= \frac{(\omega_k - \tilde{\Delta}) f_k}{h(\omega_k)} \eexp{-\ii \omega_k t}
	\equiv \alpha_k \eexp{- \ii \omega_k t}. \nonumber
\end{eqnarray}
Then, we apply Wick's theorem for example on the first element to obtain
\begin{equation}
\bra{0} b(t) b(t) A_{k_1}^\dagger A_{k_2}^\dagger \ket{0} 
	= 2 \bra{0} b(t) A_{k_1}^\dagger \ket{0} \cdot \bra{0} b(t) A_{k_2}^\dagger \ket{0} 
	= 2 \beta_{k_1} \beta_{k_2} \eexp{-\ii E_i t}
\end{equation}
where $E_i = \omega_{k_1} + \omega_{k_2}$ is the total energy. The other elements follow analogously and we obtain
\begin{equation}
\vec{v}_i
	=\left(\begin{array}{c} 
		2 \beta_{k_1} \beta_{k_2} \\ 
		\alpha_{k_1} \beta_{k_2} + \beta_{k_1} \alpha_{k_2} \\ 
		2 \alpha_{k_1} \alpha_{k_2}
	\end{array}\right)
= \left(\begin{array}{c} 
		2 \\ 
		(E- 2\tilde{\Delta})/\delta_0 \\ 
		\frac{1}{2 \delta_0^2}(E-2\dRenorm)^2 - \frac{1}{2 \delta_0^2}\epsilon^2
	\end{array}\right)
	\frac{\delta_0 f_{k_1}}{h(\omega_{k_1})} \frac{\delta_0 f_{k_2}}{h(\omega_{k_2})},
\end{equation}
where $E = \omega_{k_1} + \omega_{k_2}$ and $\epsilon = \omega_{k_1} - \omega_{k_2}$.

In addition, we need to calculate the \textbf{energy bubble}, which is given by
\begin{equation}
\mathbf{\Pi}^ \klamm{2} (z) 
= -\ii \int_{0}^{\infty} \rmd t\ \bra{0} \vec{O}_2(t) \vec{O}_2^\dagger \ket{0} \eexp{-\ii z t}
\end{equation}

By applying Wicks theorem on the four-point correlators, $\bra{0} \vec{O}_2^\dagger (t) \vec{O}_2 \ket{0}$, we obtain the product of two two-point correlators, which can expressed in terms of the first order Green's function.
As an example we calculate an element of $\mathbf{\Pi}^{(2)}(z)$, in particular,
\begin{eqnarray}
\mathbf{\Pi}^{(2)}_{11}(z)
	&=-\ii \int_{0}^{\infty} \rm d \tau \bra{0} b^2(\tau) b^{\dagger 2} \ket{0 } \eexp{-\ii z \tau} \nonumber \\
	&= -2 \ii 
	\int_{0}^{\infty} \rm d \tau 
	\int_{-\infty}^{\infty} \frac{\rm d \omega}{2 \pi} 
	\int_{-\infty}^{\infty} \frac{\rm d \omega'}{2 \pi} 
	\ii G^{(1)}_{bb} (\omega) \eexp{\ii \omega \tau} \cdot 
	\ii G^{(1)}_{bb} (\omega') \eexp{\ii \omega' \tau} \cdot
	\eexp{-\ii z \tau} \nonumber \\
	&= 2 \ii \int_{-\infty}^{\infty} \frac{\rmd \omega}{2 \pi} 
	G^{(1)}_{bb} (\omega) G^{(1)}_{bb} (z-\omega)
	= \ii (2 G^{(1)}_{bb} \ast G^{(1)}_{bb}) (z),
\end{eqnarray}
where we have introduced the convolution $(f \ast g) (E)= \int_{-\infty}^{\infty} \frac{\rmd \omega}{2 \pi} f(\omega) g(E-\omega)$.  In the discrete case, one can see that the time integral gives a $\delta$-distribution and thus the convolution, i.e., one integral and not two easily. In the continuum case, one has to argue via the poles of $h(\omega)$ and $\frac{1}{z-\omega-\omega' + \ii \eta}$.
The full self energy bubble is a 3 by 3 matrix with the elemens
\begin{equation}
\mathbf{\Pi}^{(2)}
	= \rmi \left(\begin{array}{ccc}
		2 G^{(1)}_{bb} \ast G^{(1)}_{bb} & 2 G^{(1)}_{bb} \ast G^{(1)}_{bF} & 
		2 G^{(1)}_{bF} \ast G^{(1)}_{bF}\\
		2 G^{(1)}_{bb} \ast G^{(1)}_{Fb} & 
		G^{(1)}_{bb} \ast G^{(1)}_{FF} + G^{(1)}_{bF} \ast G^{(1)}_{bF} & 
		2 G^{(1)}_{bF} \ast G^{(1)}_{FF} \\
		2 G^{(1)}_{Fb} \ast G^{(1)}_{Fb} & 2 G^{(1)}_{Fb} \ast G^{(1)}_{FF} & 
		2 G^{(1)}_{FF} \ast G^{(1)}_{FF}
	\end{array}\right).
\end{equation}

\section*{References}
\bibliographystyle{unsrt}

\end{document}